\documentclass[prb,twocolumn,showpacs]{revtex4}
\usepackage{graphicx}
\usepackage{dcolumn}
\usepackage{amsmath}

\begin{document}


\title{Structural, magnetic and electrical properties of
single crystalline La$_{1-x}$Sr$_x$MnO$_3$ for {\bf 0.4 $<$ x $<$
0.85}}

\author{J.~Hemberger$^1$, A.~Krimmel$^1$, T.~Kurz$^1$, H.-A.~Krug~von~Nidda$^1$,
V.Yu.~Ivanov$^2$, A.A.~Mukhin$^2$, A.M.~Balbashov$^3$,
A.~Loidl$^1$}
\affiliation{$^1$Experimentalphysik V, Elektronische Korrelationen
und Magnetismus,
 Institut f\"{u}r Physik,
 Universit\"{a}t Augsburg, D-86135 Augsburg, Germany \\
  $^2$General Physics Institute of the Russian Academy of
Sciences, 38 Vavilov St., 117942 Moscow, Russia \\
 $^3$Moscow Power Engineering Institute, 14 Krasnokazarmennaya St., 105835
Moscow, Russia   %
}

\begin{abstract}
We report on structural, magnetic and electrical properties of
Sr-doped LaMnO$_3$ single crystals for doping levels $0.4 \leq x
\leq 0.85$. The complex structural and magnetic phase diagram can
only be explained assuming significant contributions from the
orbital degrees of freedom. Close to $x = 0.6$ a ferromagnetic
metal is followed by an antiferromagnetic metallic phase below 200
K. This antiferromagnetic metallic phase exists in a monoclinic
crystallographic structure. Following theoretical predictions this
metallic antiferromagnet is expected to reveal an
(x$^2$-y$^2$)-type orbital order. For higher Sr concentrations an
antiferromagnetic insulator is established below room temperature.

\end{abstract}

\pacs{75.30.-m, 77.30.Kg, 72.80.Ga }

\maketitle


\section{Introduction}
The fascinating phase diagrams of the doped maganites result from
a subtle interplay of spin, charge, orbital and lattice degrees of
freedom. In La$_{1-x}$Sr$_x$MnO$_3$ the main body of experimental
investigations has been carried out for Sr concentrations  $x <
0.5$. This partly has been due to the fact that colossal
magnetoresistance effects \cite{vonHelm93_Chah93} show up around
$x=0.3$, which in the beginning of the research seemed to be
rather promising for application. On the other hand single
crystals for $x > 0.5$ are hard to grow and hence, only rarely
have been investigated.

Already at low concentrations ($x < 0.5$) a rather complex phase
diagram evolves, 
which is due to the fact that in addition to super-exchange (SE)
and double-exchange (DE) interactions, charge order (CO) and
structural effects via orbital ordering are of outstanding
importance. For $x < 0.2$, a long-range cooperative Jahn-Teller
(JT) effect establishes orbital order (OO), which finally
determines the antiferromagnetic (AFM) spin state of A-type in the
pure compound at low temperatures. With increasing Sr doping $(x <
0.1)$ a ferromagnetic (FM) component evolves in addition to the
AFM order of subsequent planes which is explained in terms of
electronic phase separation \cite{Dagot01} or, following the
time-honored ideas of de Gennes \cite{deGen60}, in terms of a
canted AFM (CA) state. However, within this model of competing SE
and DE interactions clearly the importance of lattice distortions
has to be taken into account.\cite{Millis} On further increasing
$x$ $(0.1 \leq x \leq 0.17)$, it seems clear that a new type of
orbital order, probably connected with CO, determines the
low-temperature insulating ferromagnet (FM) around $x =
0.125$.\cite{Endoh99,Parask00,Niem99} Finally, for Sr
concentrations $x
> 0.17$ the long-range JT distortions become suppressed and a
ferromagnetic metal evolves below the FM phase transition which is
stable almost up to half filling.

After the early work on La:SrMnO$_3$ which is summarized by
Goodenough and Longo\cite{Goode70}, the complex phase diagram has
been studied by many groups and a more or less coherent picture
can be deduced from the variety of experimental results
reported.\cite{Parask00,Urus95,Yama96,Kawa96,Zhou97} From the
systems with narrower bandwidth (e.g. $L_{1-x}A_x$MnO$_3$ with $L$
= Nd, Pr, Sm and $A$ = Ca, Sr \cite{Maez98,Mart99}) an extreme
asymmetry between the hole and electron doped regimes is well
known and it seems highly interesting also to investigate the
electron doped La$_{1-x}$Sr$_x$MnO$_3$. However, for $x
> 0.5$ much less experimental information is available.

Structural, resistivity and magnetization results for $x = 0.5$
and 0.54 were reported by Akimoto et al. \cite{Akim98} For $x =
0.54$ they reported an orthorhombic ($Pbnm$) nuclear and an A-type
AFM structure at 10 K. Results for similar Sr concentrations $0.5
< x < 0.6$ were published by Moritomo et al. \cite{Morit98} On
increasing $x$, the crystal symmetry changes from rhombohedral
($R3c$) to pseudotetragonal  at $x = 0.54$ with $a \approx b <
c\sqrt{2}$. For concentrations around $x = 0.55$, a metallic AFM
phase, with A-type spin structure at low temperatures, is followed
by a FM metallic (M) state at elevated temperatures.\cite{Morit98}
Polycrystalline La$_{1-x}$Sr$_x$MnO$_3$ has been investigated by
Fujishiro et al. \cite{Fuji98} by magnetization, electrical
resistivity and ultrasonic techniques. They arrive at a different
phase diagram, with an insulating (I) state for all concentrations
$x > 0.5$. Similar findings were reported by Patil et al.
\cite{Patil00} who also investigated ceramic samples for $0.46
\leq x \leq 0.53$. They found a sequence of magnetic and
charge-order transitions and interpreted their results in terms of
electronic phase separation.

Further interest in the overdoped manganites arises from the fact
that for the insulating regions of the phase diagram electronic
phase separation in form of stripes \cite{Rada99} or bistripes
\cite{Mori98} has been reported. This special form of CO certainly
is driven by OO.\cite{Khom01} Also theoretically the phase
diagrams of the doped manganites have been investigated in great
detail.\cite{Dagot01,Maez98,Khom01,Held98,Brink99} Specifically in
Ref. \cite{Brink99} the overdoped regime was investigated using DE
within degenerate orbitals. Depending on the bandwidth, on
increasing electron doping a sequence of spin structures of type
A, C, A and F has been predicted.\cite{Brink99,Good63}

In order to clarify the situation and to shed some light on the
complex and complete ($x,T$)-phase diagram in
La$_{1-x}$Sr$_x$MnO$_3$ we grew a series of single crystals for
concentrations ($0.4 < x < 0.85$). We were not able to grow
crystals with higher Sr concentrations. The reason seems to be
rather clear as SrMnO$_3$ reveals a hexagonal crystal structure
and obviously there exists a miscibility gap for concentrations
close to pure SrMnO$_3$. In the following we present detailed
structural, magnetic susceptibility, magnetization and electrical
resistivity results for the complete series of crystals. From the
results we construct a detailed phase diagram. Polycrystalline
SrMnO$_3$ has been investigated to complete the phase diagram.
These experiments are a continuation of earlier work on crystals
with low Sr doping levels which has been published previously
\cite{Parask00} and has been included to present the complete
($x,T$)-phase diagram of La$_{1-x}$Sr$_x$MnO$_3$ for $0 \leq x
\leq 1$.

\section{Experimental Details}

La$_{1-x}$Sr$_x$MnO$_3$ single crystals were grown by a
floating-zone method with radiation heating similar to the
techniques as described in Ref.\cite{Balba96} For these crystals
with Sr concentrations $x \geq 0.5$, different atmospheric
conditions ranging from air atmosphere to excess oxygen pressure
up to 50 atm were tested to optimize the growth process.
Nevertheless an uncertainty in the control of the concentration
$x$ of up to 2\% cannot be ruled out. To complement the phase
diagram pure SrMnO$_3$ has been grown using standard ceramic
techniques. Powder x-ray diffraction measurements were performed
utilizing Cu-K$_{\alpha}$ radiation with $\lambda = 0.1541$~nm.

The magnetic susceptibility and magnetization were measured using
a commercial superconducting quantum interference device (SQUID)
magnetometer ($1.5 < T < 400$~K, $H\leq50$~kOe) and an
ac-magnetometer which operates up to magnetic fields of $H =
140$~kOe. The electrical resistance has been measured using
standard four-probe techniques in home-built cryostats and ovens
from 1.5~K to 600~K.

\begin{figure}[htbp]
\includegraphics[clip,width=80mm]{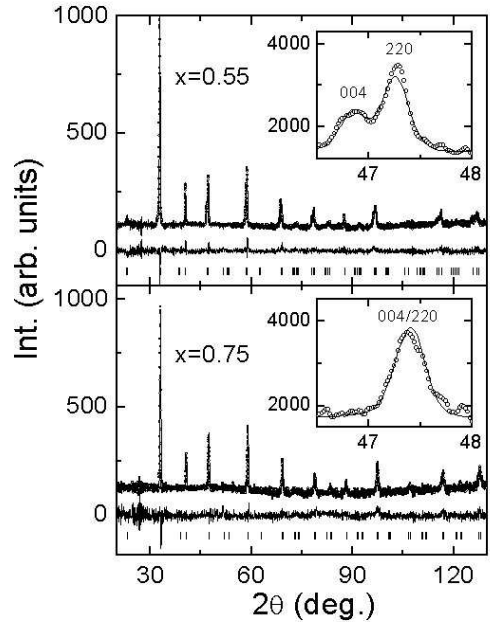}
\caption{\label{xraydiff}   %
X-ray diffraction profiles of La$_{1-x}$Sr$_x$MnO$_3$ for
concentrations $x = 0.55$ (upper frame) and $x = 0.75$ (lower
frame). The solid lines correspond to the results of a Riet\-veld
refinement. The difference patterns are indicated in each frame.
The insets show the splitting of the [220] and [004] reflexes,
denoting the decrease of the tetragonal
distortion with increasing $x$.%
}%
\end{figure}
\begin{figure}[htbp]
\includegraphics[clip,width=75mm]{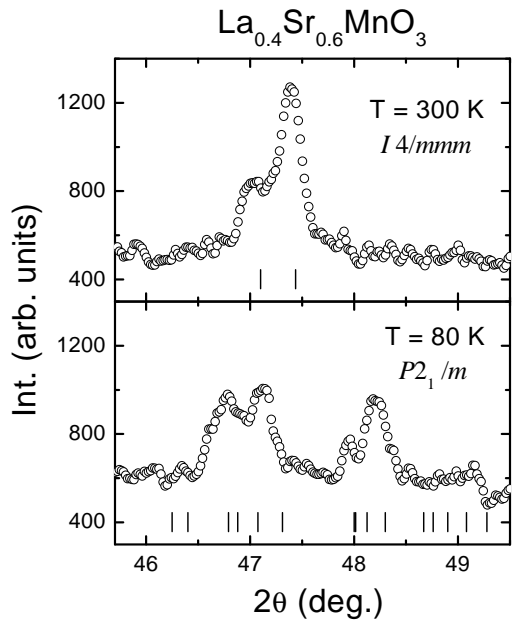}
\caption{\label{diff2}   %
Parts of the temperature dependent x-ray powder-diffraction
patterns of La$_{0.4}$Sr$_{0.6}$MnO$_3$ for
$46^\circ<2\Theta<49^\circ$. At $T=80~K$ pronounced superlattice
reflections around $2\Theta\approx48^\circ$ indicate a
monoclinic distortion.%
}%
\end{figure}

\begin{table*}[htbp]
\caption{
Room temperature crystal symmetry and lattice parameters of La$_{1-x}$Sr$_{x}$MnO$%
_{3}$ for Sr-concentrations ($0.5\leq x\leq 1.0$). The diffraction
data for $x<1$ were derived from single crystalline material which
was powdered for the diffraction measurements. The two lowest rows
display Curie-Weiss temperatures $T_{CW}$ and the effective
paramagnetic moments $\mu_{eff}$ obtained from the linear regime
of the inverse susceptibility below $T=400$~K.
}
\begin{ruledtabular}
\begin{tabular}{c | c c c c c c c }
$x$ & 0.5 & 0.55 & 0.6 & 0.65 &  0.75 & 0.85 & 1.0 \\
\hline
\begin{tabular}{l}
crystal \\
structure
\end{tabular}
&
\begin{tabular}{l}
rhombo- \\
hedral
\end{tabular}
&
\begin{tabular}{l}
tetra-\\ gonal
\end{tabular}
&
\begin{tabular}{l}
tetra- \\ gonal
\end{tabular}
&
\begin{tabular}{l}
tetra- \\ gonal
\end{tabular}
&
\begin{tabular}{l}
tetra- \\ gonal
\end{tabular}
&
\begin{tabular}{l}
tetra- \\ gonal
\end{tabular}
&
\begin{tabular}{l}
hexa-\\ gonal
\end{tabular}
\\
$a$ [\AA ] & 5.461 & 5.438 & 5.448 & 5.437 & 5.419 & 5.404 & 5.452 \\
$c$ [\AA ] & - & 7.753 & 7.672 & 7.670 & 7.686 & 7.665 &  9.084  \\
$c/\sqrt{2}a$  & - & 1.008 & 0.996 & 0.997 & 1.002 & 1.003 &  -  \\
$\alpha $ [$^{\circ }$] & 60.16 &  &  &  &  &  &  \\
$T_{CW} $ [K] & 333 & 313 & 299 & 255 & 64 & -2 & -980 \\
$\mu_{eff} $ [$\mu_B$] & 5.1 & 5.0 & 4.9 & 4.8 & 4.9 & 4.6 & 4.0 \\
\end{tabular}
\end{ruledtabular}
\end{table*}

\section{Results and discussion}
\subsection{X-ray diffraction}

To demonstrate the quality of the single crystals under
investigation, we powdered pieces of the single crystals used for
magnetic and transport measurements and performed detailed
Rietveld refinements of the diffraction profiles for samples with
Sr concentrations $0.55\leq x \leq 0.85$ at room temperature.
Fig.~\ref{xraydiff} shows the diffraction profiles, the refinement
and the difference pattern for $x = 0.55$ (upper panel) and $x =
0.75$ (lower panel). First of all we want to stress that all
reflections can be indexed and no impurity phases are apparent
above the background level, even no spurious amount of SrMnO$_3$
which has been reported in all previous
investigations.\cite{Goode70,Akim98,Morit98,Fuji98} At first sight
one recognizes that with increasing Sr concentration $x$ the
crystals almost approach cubic symmetry which is expected for
concentrations with a tolerance factor close to 1. This is
demonstrated by the insets in Fig.~\ref{xraydiff}, showing the
splitting of the 220/004 reflections which provides direct
experimental evidence of the $c/a$ ratio and hence on the
tetragonal distortion. For $x = 0.55$ a clear splitting of the
reflections can be detected, while for $x = 0.75$ no apparent
splitting is visible and a splitting can only be derived via the
broadening of the reflection in a detailed Rietveld refinement
with well defined resolution parameters. Hence the samples near $x
= 0.8$ are very close to cubic symmetry.

From the Rietfeld refinement we determined the lattice constants
of the samples under investigation. The results are listed in
Tab.~1. At room temperature the structure changes from
rhombohedral (R) at $x = 0.5$ to tetragonal (T) at $x = 0.55$ and
finally to hexagonal (H) close to pure SrMnO$_3$. We would like to
recall that SrMnO$_3$ was only prepared in ceramic form and we
were not able to grow crystals beyond strontium concentrations $x
= 0.85$. The room-temperature tetragonal phase extends over a
broad concentration  range, with a significant change of the
tetragonal distortion.  For $x = 0.55$ we find a value of
$c/\sqrt{2}a \approx 1.01$. Note that for this concentration
ferromagnetism is established already at room temperature. On
increasing $x$ this ratio is reduced yielding values close to 1,
however now in the paramagnetic phase, but still on the verge of
magnetic order.
It is clear that the $c/a$ ratio will strongly depend on the
orbital structure, i.e. whether
the orbitals are aligned as $d_{x^2-y^2}$ or as $d_{3z^2-r^2}$.
However, systematic temperature dependent x-ray diffraction
measurements are needed to arrive at final conclusions. From
preliminary temperature-dependent x-ray diffraction experiments
(down to 80~K) we found that for $0.5 < x < 0.7$ the symmetry is
lowered and the structure changes into a monoclinic (Mc) phase
characterized by space group $P2_1/m$. Fig.~\ref{diff2}
illustrates the evolution of superlattice reflections for $x=0.6$
due to the monoclinic distortion on cooling. At the boundaries of
this monoclinic low-temperature phase for $x\approx 0.5$ and
$x\approx 0.7$ we suggest that the system enters into a
structurally mixed phase where the Mc phase coexists with a R and
T phase, respectively. However a more detailed evaluation is
necessary to analyze these two-phase regions. A detailed report on
the structural properties will be given elsewhere.\cite{Krim02} It
should be noticed that a low-temperature monoclinic phase with
space group $P2_1/m$ has been established in the doped manganites
La$_{0.5}$Ca$_{0.5}$MnO$_3$ and
Pr$_{0.6}$(Ca$_{1-x}$Sr$_x$)$_{0.4}$MnO$_3$, as well as in
Pr$_{0.5}$Sr$_{0.5}$MnO$_3$.\cite{Rad97,Lee98, Kawa97}
Furthermore, the presence of monoclinic domains within an
orthorombic matrix has been reported for
Pr$_{0.7}$Ca$_{0.25}$Sr$_{0.05}$MnO$_3$ and
Pr$_{0.75}$Sr$_{0.25}$MnO$_3$ on the basis of high-resolution
microscopy.\cite{Her96}

\subsection{Transport properties}

\begin{figure}[htbp]
\includegraphics[clip,width=70mm]{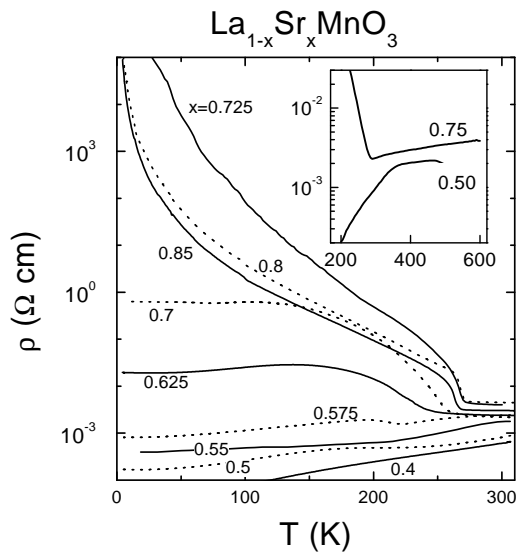}
\caption{\label{resistivity}  %
Temperature dependence of the resistivity in
La$_{1-x}$Sr$_x$MnO$_3$ for different concentrations $0.4 \leq x
\leq 0.85$ as indicated in the figure. The inset shows the
resistivity
for $x = 0.5$ and $x = 0.75$ measured also at higher temperatures. %
}%
\end{figure}

Fig.~\ref{resistivity} shows the temperature dependence of the
electrical resistivity below room temperature. Clearly
La$_{1-x}$Sr$_x$MnO$_3$ reveals a metallic conductivity for $0.4 <
x < 0.6$. These findings differ from those reported by Patil et
al. \cite{Patil00} for concentrations close to $x=0.5$ obtained
from ceramic samples, where the resistivity increases towards low
temperatures. From optical measurements these authors find
indications for an anisotropy of the conductivity. This fact
together with possible grain effects in ceramic samples could
explain the different results.

For Sr concentrations $x = 0.625$ and 0.7 a significant increase
of the resistivity occurs below 250~K. Finally in the samples with
$x > 0.7$ this feature is sharper and denotes the transition from
the paramagnetic to the antiferromagnetic state.
Towards low temperatures a purely semiconducting behavior can be
found. It seems that the samples with Sr concentration $x \approx
0.7$ belong to a two-phase region separating metallic and
insulating regimes. At the same time the system is close to a
phase boundary between the monoclinic and tetragonal crystal
structures.
Nevertheless, the sample with $x=0.7$ remains a candidate for the
occurrence of electronic phase separation for $T<200$~K. To
investigate also the high-temperature electronic behavior, we
measured the electrical resistance for some representative samples
up to 600~K. The results for $x = 0.5$ and $0.75$ are shown in the
inset. For $x = 0.5$ the metallic behavior at low temperatures
changes into the temperature characteristics of a bad metal at
elevated temperatures. %
For $x > 0.7$ a bad metallic behavior for $T > 250$~K changes into
strongly semiconducting temperature characteristics below. The
low-temperature resistivity as a function of Sr concentration
exhibits a maximum around $x\approx0.75$. This may indicate a CO
state at electronic quarter filling of the $e_g$ bands.

\subsection{Magnetization and magnetic susceptibility}

\begin{figure}[bhtp]
\includegraphics[clip,width=80mm]{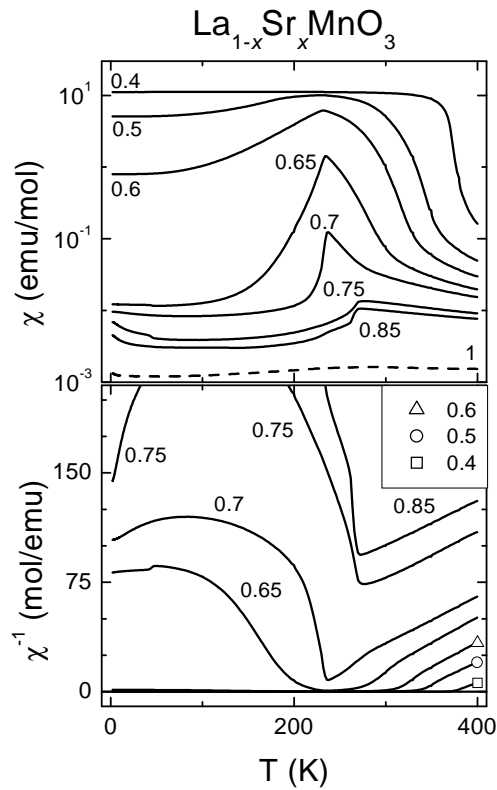}
\caption{\label{suscep}  %
Temperature dependence of the magnetic susceptibility in
La$_{1-x}$Sr$_x$MnO$_3$ as measured at a dc magnetic field of
1~kOe, for Sr concentrations from 0.4 to 0.85. In the upper frame
the susceptibility is shown as a function of temperature. The
lower frame shows the inverse susceptibilities vs. $T$. %
}%
\end{figure}

The magnetic dc-susceptibility (M/H measured at 1 kOe) of
La$_{1-x}$Sr$_x$MnO$_3$ is shown in Fig.~\ref{suscep} in the upper
frame. For $x = 0.4$, at 370~K we find the pure ferromagnetic
phase transition, which is driven by DE interactions and is
characteristic for the materials showing a colossal
magneto-resistance (CMR) effect. For $x = 0.5$ and 0.6 the
magnetization decreases below 250~K indicating that the
ferromagnetic moments become reduced at low temperatures. This may
be due to a slight canting of the spins or due to electronic
phase-separation effects. Similar findings for $0.46\leq x \leq
0.53$ were interpreted in terms of phase separation within a
charge ordered state.\cite{Patil00} We would like to recall that
for $x\approx0.5$ we found indications of a two-phase region of a
rhombohedral and a monoclinic structure. At the moment it is
unclear, whether this is a structural two-phase region, or whether
the groundstate can be explained in terms of electronic phase
separation between FM and AFM regions including scenarios like
stripe or bistripe formation.

Well defined cusps appear for $x = 0.65$ and 0.7 which indicate
the vanishing of a spontaneous ferromagnetic moment and the
presence of a nearly completely antiferromagnetic (AFM)
spin-structure at low temperatures. It is important to note that
we observed a pronounced temperature hysteresis at these cusps
typical for a first-order magnetic transition. Finally for $x
> 0.7$ the characteristics of purely antiferromagnetic phase
transitions are detected. From the inverse susceptibilities, which
are shown in the lower frame of Fig.~\ref{suscep}, we can deduce
the Curie-Weiss (CW) temperatures, which continuously decrease
while the Sr concentration increases from $x = 0.4$ to $x = 0.7$
(cf. Tab.~1). Beyond this concentration a significant decrease of
the CW temperatures appears and finally, for $x =0.85$, a negative
CW temperature can be read off. For the concentrations $0.55 \leq
x \leq 0.7$ in addition a distinct deviation from the CW behavior
can be detected well above the magnetic ordering temperature
indicating the presence of strong spin fluctuations or even of
short-range magnetic order.

\begin{figure}[htbp]
\includegraphics[clip,width=80mm]{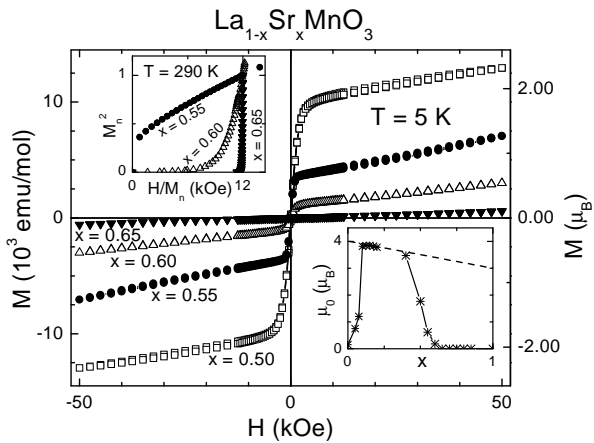}
\caption{\label{xhys} %
Low-temperature ($T = 5$~K) magnetization of
La$_{1-x}$Sr$_x$MnO$_3$ as a function of magnetic field for
concentrations $x = 0.5$, 0.55, 0.6  and 0.65. The lower right
inset shows the remnant (ferromagnetic) saturated moment as a
function of concentration. The dashed line indicates the saturated
moment assuming that all Mn ions (Mn$^{3+}$ and Mn$^{4+}$)
contribute only with their spin values to the ordered moment. The
upper left inset shows the normalized magnetization at 290~K for
concentrations $x = 0.5$, 0.55, and 0.6 presented as $M_n^2$ vs.
$H/M_n$ with $M_n=M/M(12$kOe$)$.
}%
\end{figure}

To further elucidate the low-temperature magnetic properties of
La$_{1-x}$Sr$_x$MnO$_3$, Fig.~\ref{xhys} shows the magnetization
at 5~K for a series of crystals with concentrations $0.5 < x <
0.65$. An almost pure antiferromagnet at $x = 0.65$ is followed by
an increasing ferromagnetic component on decreasing $x$. But even
at $x = 0.5$ only 2/3 of the full possible ferromagnetic
magnetization is observed at low fields, and on increasing
external field the magnetization increases strictly linearly,
which can be explained in terms of a canted antiferromagnet, whose
canting angle becomes reduced in increasing fields, or in terms of
electronic phase separation. Including earlier published
results\cite{Parask00}, the lower inset of Fig.~\ref{xhys}
demonstrates, how the ferromagnetic moment evolves in the complete
concentration regime. The expected spin-only FM moment only
evolves for Sr concentrations $0.2 < x < 0.4$ and approaches
values close to zero for lower and higher Sr concentrations. The
upper inset of Fig.~\ref{xhys} compares the field dependence of
the magnetization for the concentrations $x=0.55$, 0.6, and 0.65
at $T=290$~K. The representation $M_n^2$ vs. $H/M_n$
(Arrot-plot\cite{Arro67}) demonstrates the presence of a
spontaneous FM component (i.e. finite $M^2$ at $H=0$) for
$x=0.55$. At the same temperature the behavior of $x=0.65$ is
purely paramagnetic (i.e. $H/M$ is constant). The curve for
$x=0.6$ exhibits an intermediate behavior, denoting the presence
of strong FM fluctuations or short-range order (SR). This
corresponds to the strong deviations from the Curie-Weiss behavior
discussed earlier (see Fig.~\ref{suscep}).

\begin{figure}[htbp]
\includegraphics[angle=0,width=80mm,clip]{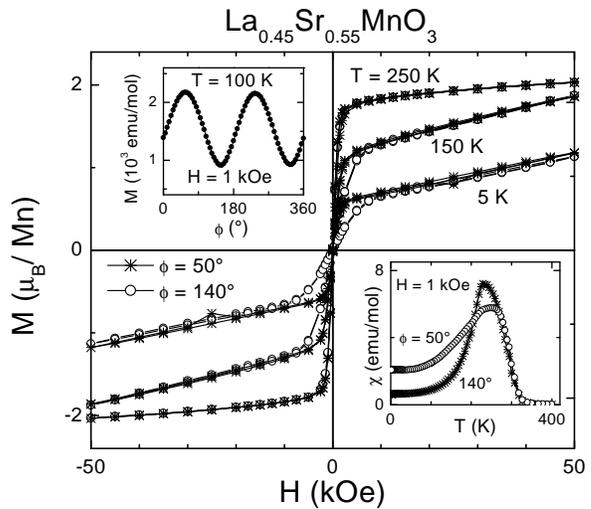}
\caption{\label{thys}  %
Magnetization as a function of external field in
La$_{0.45}$Sr$_{0.55}$MnO$_3$ measured at different temperatures
and different orientations with respect to the external field (see
upper inset). The solid lines are drawn to guide the eye.
The lower inset displays the temperature dependence of the
magnetic dc susceptibility $\chi=M/H$, measured at these different
orientations.
}%
\end{figure}

In what follows, we wanted to study the magnetic anisotropy close
to half filling in more detail.
Around $x = 0.55$ the ferromagnetic phase at elevated temperatures
is followed by a further magnetic phase transition which
significantly reduces the magnetization (see Fig.~\ref{suscep}).
To elucidate the temperature evolution of this weak ferromagnetic
regime in more detail, Fig.~\ref{thys} shows the magnetization vs.
external field in La$_{0.45}$Sr$_{0.55}$MnO$_3$ for a series of
temperatures. At 250 K, just below the FM phase transition a
strong ferromagnetic hysteresis evolves and, taking into account
the elevated temperature, almost the full saturated moment of
Mn$^{3+}$/Mn$^{4+}$ can be detected. However, below 200~K the
magnetization at moderate fields becomes significantly reduced and
reveals a strictly linear increase of $M$ as function of field.
Finally at 5~K the FM component amounts approximately 0.5 $\mu_B$
only.

To study the anisotropy of this magnetic ground state, the data
were taken for two different orientations of the sample with
respect to the external field. The upper inset of Fig.~\ref{thys}
shows the angular dependence of the magnetization as observed at
$T=100$~K when the sample was rotated around an axis approximately
within the $a,b$-plane. From this a hard and an easy axis were
defined, which differ by almost a factor of 2.5 in the
magnetization values. The lower inset of Fig.~\ref{thys} shows the
temperature dependence of the magnetic susceptibility measured for
these both orientations. The distinct evolution of anisotropic
behavior below $T\approx250$~K can be observed. This behavior
clearly is different from the anisotropy observed for strontium
concentrations of $x = 0.05$ \hspace{0.1cm} where the
magnetization is zero along $a$ and $b$ and finite along $c$,
indicating a slight canting of the moments out of the
$a,b$-plane.\cite{Parask00} It seems that for $x = 0.55$ the AFM
structure at low temperatures is certainly more complex.
Nevertheless, it has to be stated, that neither a considerable
uncertainty concerning the determination of the $a,b$-plane nor
possible twinning of the sample can be ruled out. However, as
observed for $x = 0.05$ \hspace{0.1cm}\cite{Parask00} the
paramagnetic susceptibility again is fully isotropic for
temperatures above the magnetically ordered regime. It is
interesting to note that the anisotropy changes in the (canted)
antiferromagnetic state. These results indicate either a canted
spin structure or phase separation assuming ferromagnetic clusters
within an antiferromagnetic matrix. But, of course, we cannot
exclude more complex phase-separation
scenarios.\cite{Rada99,Mori98} It also should be kept in mind that
the sample under consideration still behaves metallic, in the
paramagnetic (PM), the FM, and the canted AFM state, respectively.

\subsection{Phase diagram}

\begin{figure}[htbp]
\includegraphics[angle=-90,clip,width=80mm]{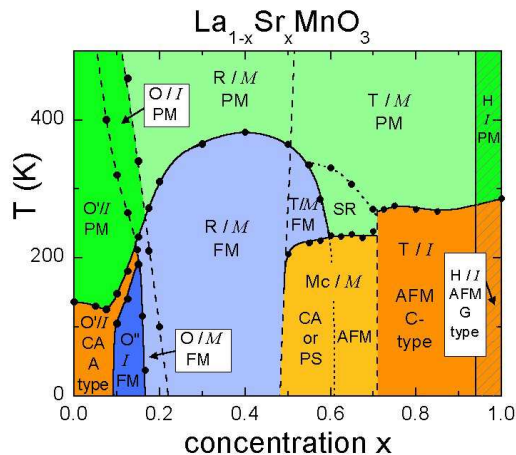}
\caption{\label{phasediag} %
Phase diagram of La$_{1-x}$Sr$_x$MnO$_3$ for the complete
concentration regime, including results of Ref.\cite{Parask00} The
crystal structures (Jahn-Teller distorted orthorhombic: O',
orthorhombic O; orbital-ordered orthorhombic: O'', rhombohedral:
R, tetragonal: T, monoclinic: Mc, and hexagonal: H) are indicated
as well as the magnetic structures (paramagnetic: PM (green),
short-range order (SR), canted (CA), A-type antiferromagnetic
structure: AFM (yellow); ferromagnetic: FM (blue), phase separated
(PS), and AFM C-type structure) and
the electronic state (insulating: I (dark); metallic: M (light)).%
}%
\end{figure}

Finally in Fig.~\ref{phasediag} we present the phase diagram for
the complete concentration regime $0 \leq x \leq 0.85$ where mixed
single crystals can be grown. We included the results from earlier
work \cite{Parask00} ($0 \leq x \leq 0.3$). The complex sequence
of magnetic phases at low concentrations is highly influenced by
the cooperative JT distortion of the O' phase and by the orbital
order of the O'' phase.\cite{Endoh99} For $x > 0.2$ a large
ferromagnetic regime evolves which reveals a rhombohedral
structure and shows CMR effects throughout. For $x > 0.5$ a
tetragonal phase appears which is a FM metal. On decreasing
temperature it undergoes a transition into a monoclinic
antiferromagnetic state, but still exhibits metallic behavior. We
would like to recall that in Pr$_{0.5}$Sr$_{0.5}$MnO$_{3}$ and
Nd$_{0.5}$Sr$_{0.5}$MnO$_{3}$ the FM and M phase is followed by an
insulating AFM phase, which reveals a CE-type spin structure for
the Nd and an A-type layered spin structure for the Pr
compound.\cite{Kawa97} The latter shows no clear signs of
CO.\cite{Kawa97} However, for Nd$_{0.45}$Sr$_{0.55}$MnO$_{3}$ a
metallic behavior has been detected within the ferromagnetic
layers, while the resistivity along c revealed a semiconducting
characteristic.\cite{Kuwa99} It seems that in the La:Sr series of
the manganites this two dimensional metallic phase, which is
antiferromagnetic, evolves for $x \approx 0.6$. There the
resistivity remains fully metallic ($d\rho/dT>0$) for all
temperatures (see Fig.~\ref{resistivity}) and at $T=5$~K only a
weak ferromagnetic moment can be detected (Fig.~\ref{xhys}).

This metallic and AFM state is monoclinic and is embedded into
structurally different regions (R for $x<0.5$ and T for $x>0.7$),
depicted schematically by vertical dashed lines in
Fig.~\ref{phasediag}. It seems naturally to assume that this
exotic phase also is characterized by a new orbital order. Close
to $x = 0.5$ it seems that the M/FM phase coexists with the M/AFM
phase. According to preliminary diffraction experiments indeed
also the crystallographic phases coexist. We would like to recall
that from observations in conventional ferromagnets a variety of
domain structures can be formed in multicomponent systems,
including cluster or stripe formation. Whether in this case a
description in terms of electronic phase separation is adequate,
has to be proven in detailed experiments. At present it is also
unclear how the observation of stripes \cite{Rada99} or bistripes
\cite{Mori98} is compatible with the observed magnetic anisotropy.
For concentrations $x \approx 0.7$ metallic and insulating domains
cannot be ruled out. This we conclude from the temperature
dependence of the resistivity. At the moment it remains an open
question, whether a T/I/AFM phase coexists with an Mc/M/AFM state
or whether in this region electronic phase separation evolves out
of the structurally pure phase.

A further interesting phenomenon in this concentration range is
the fact, that the AFM metallic phase evolves out of a
ferromagnetic metal. Probably double-exchange interactions drive
the ferromagnet, while superexchange interactions are responsible
for the AFM state. This can only be explained, if the orbital
order changes as a function of temperature.

For concentrations $x > 0.75$ a purely AFM and insulating state
evolves within a nearly cubic structure. For these concentrations
the tolerance factor is close to 1. For further increasing
concentrations the ABO$_3$ perovskite structure is unstable and no
mixed crystals can be grown. SrMnO$_{3-\delta}$ reveals a layered
perovskite structure.

\section{Conclusion}

What did we learn from these experiments in addition to the
existing enormous amount of knowledge on the doped manganites? To
summarize, we were able to grow high-quality single crystals of
La$_{1-x}$Sr$_x$MnO$_3$  for $0 \leq x \leq 0.85$ with no
parasitic phases like SrMnO$_3$. We constructed a rather complete
($x,T$) phase diagram (Fig.~\ref{phasediag}), which reveals the
well established asymmetry between the hole doped $(x < 0.5)$ and
the electron doped $(x > 0.5)$ regime. Of course, an essential
part of this asymmetry is driven by geometrical constraints via
the tolerance factor, which increases from $t = 0.89$ for $x = 0$
to $t = 1.01$ for $x = 1$. Hence, on increasing concentration,
La$_{1-x}$Sr$_x$MnO$_3$ reveals a decreasing buckling of the
MnO$_6$ octahedra. But partly this asymmetry also is driven by
orbital degeneracy. Close to $x = 0$ strong JT distortions reveal
orbital order which concomitantly determines the A-type spin
structure. For higher Sr concentrations and in the crystals with a
high symmetry at the Mn site, orbital degeneracy will play an
essential role, yielding completely different spin ground
states.

In Fig.~\ref{phasediag} the combined influence of the
concentration dependence of the tolerance factor and the
increasing importance of orbital degeneracy as $x$ increases is
nicely documented in the sequence of structural phases: At room
temperature and for $0 \leq x \leq 0.85$ we find the
crystallographic phases O' orthorhombic, O orthorhombic,
rhombohedral and tetragonal. The corresponding series of
electronic ground-state properties is insulating/spin A-type AFM,
insulating/spin FM, metallic/spin FM, metallic/spin A-type AFM;
insulating/spin C-type AFM, and finally insulating/spin G-type for
$x =1$. So far we did not investigate the spin structures of the
magnetic phases for $x \geq 1$.
Theoretically, taking SE, DE and orbital degeneracy into account
the sequence of spin structures A-F-A-C-G has been
predicted.\cite{Maez98} We want to stress that a pure insulating
AFM state appears in all phase diagrams at high doping levels and
that a G-type antiferromagnet is stable for $x = 1$. The spin
structure close to $x=0.8$ (AFM, C-type) has been taken in analogy
to other doped manganites and to the existing literature.
For this concentration regime around $x=0.8$ evidence for charge
order can be found due to the sharp increase of the resistivity at
$T_N$, possibly in the form of stripe or bistripe structures as it
has been suggested in literature.\cite{Mori98} Close to $x=0.6$ it
has been proposed that the orbitals are strictly aligned within
the $a,b$-plane forcing an A-type spin structure.\cite{Maez98}
La$_{0.4}$Sr$_{0.6}$MnO$_3$ reveals antiferromagnetic order with a
FM component, which is of the order of $0.1~\mu_B$ at low fields.
It is this concentration regime where also two-dimensional
metallic behavior is expected.

In what follows we would like to compare the phase diagram of
La$_{1-x}$Sr$_{x}$MnO$_3$  as observed in the present
investigations with published phase diagrams of other manganites
and with theoretical predictions. This discussion is based on
experiments covering the hole and electron doped regime of the
phase diagram and investigating the half-doped case in detail.
Experimental phase diagrams are available for La:Sr \cite{Niem99},
La:Ca \cite{Schiff95} , Pr:Sr \cite{Maez98,Mart99,Kawa97}, Pr:Ca
\cite{Mart99,Tomi96}, Nd:Sr \cite{Maez98,Kawa97}, Sm:Sr and Sm:Ca
\cite{Mart99} manganite. From these results a generic electronic
phase diagram can be deduced, which yields the sequence of AFM
(A-type)/I, FM/I, FM/M and AFM/I (C-type). The latter spin
structure dominates a broad region in the electron doped ($x >
0.5$) crystals. It is clear that these spin structures and
electronic properties are closely linked with the orbital degrees
of freedom, and the orbital order for most of these phases has
been theoretically proposed by Maezono et al. \cite{Maez98},
Khomskii \cite{Khom01}, van den Brink et al. \cite{Brin99} and by
Solovyev and Terakura.\cite{Solo99}

It is worth mentioning that there are some noticeable exceptions
from this universal phase diagram: e.g. the Sm:Ca \cite{Mart99}
and the Pr:Ca \cite{Mart99,Tomi96} compounds do not exhibit a
typical CMR regime with a ferromagnetic and insulating ground
state for $x < 0.5$. This fact probably results from geometrical
constraints. In addition peculiarities show up close to half
filling: In the Nd:Sr and Pr:Sr compounds close to $x = 0.5$
charge ordered antiferromagnetic phases were found.\cite{Kawa97}
Magnetically these systems form ferromagnetic zig-zag chains which
are coupled antiferromagnetically. This CE-type of magnetic
structure results from charge and orbital order and has been
explained theoretically in detail.\cite{Solo99,Brin99} The FM
zig-zag chains with their concomitant charge order can easily be
molten by the application of moderate magnetic
fields.\cite{Kuwa95}  In addition, van den Brink et al.
\cite{Brin99} have shown that for doping $x < 0.5$ the CE
structure is unstable against phase separation. This CE-type spin
structure with a checkerboard-type CO is not observed in the La:Sr
compound under investigation. In our high-quality samples we find
an AFM and metallic ground state with no signs of charge order.
Hence we do not expect to find a CE-type magnetic structure. A
metallic AFM spin A-type phase has been predicted by Maezono et
al.\cite{Maez98} which reveals an (x$^2$ - y$^2$) orbital
structure. In this structure hopping along c is forbidden and the
metallic conductivity is strictly two dimensional and indeed this
type of behavior has been observed in
Nd$_{0.45}$Sr$_{0.55}$MnO$_{3}$.\cite{Kuwa99} We believe that this
type of spin and orbital structure may also be present in
La$_{1-x}$Sr$_{x}$MnO$_3$ for  $x \approx 0.6$. Measurements of
the magnetic structure and of the anisotropy of the electronic
transport still have to be performed. It remains a puzzle why in
Nd$_{0.45}$Sr$_{0.55}$MnO$_3$, which reveals a metallic, spin
A-type phase, Kuwahara et al.\cite{Kuwa99} observed a significant
anisotropy in the resistivity but not in the magnetic
susceptibility. In contrast La$_{0.45}$Sr$_{0.55}$MnO$_3$ exhibits
a pronounced anisotropy both in the FM and in the AFM states (cf.
Fig.~6).
It seems clear that the FM metallic state at elevated temperatures
still is driven by double exchange, and perhaps the
low-temperature metallic state indicates electronic phase
separation where the FM metallic paths still percolate. The
occurrence of a FM metal followed by an AFM metallic state for
concentrations close to $x=0.6$ still is a puzzle and further
experimental work is needed to elucidate this strange phase.

\acknowledgments This work was supported by the Bundesministerium
f\"{u}r Bildung und Forschung (BMBF) via VDI/EKM, FKZ 13N6917, by
the Deutsche Forschungsgemeinschaft via SFB 484 (Augsburg), and by
INTAS via project 97-30850.


\bibliographystyle{prsty}
\bibliography{latio3,labatio3}




\end{document}